\documentclass[%
 reprint,
 superscriptaddress,
amsmath,amssymb,
prl,
floatfix,
]{revtex4-2}

\usepackage{float}
\usepackage{graphicx}
\usepackage{dcolumn}
\usepackage{bm}
\usepackage{xcolor}
\usepackage[version=4]{mhchem} 

\graphicspath{{Figures/}}

\begin{document}

\preprint{APS/123-QED}

\title{Twisto-electrochemical activity volcanoes in Trilayer Graphene}

\author{Mohammad Babar}
\affiliation{Department of Mechanical Engineering, Carnegie Mellon University, Pittsburgh, Pennsylvania 15213, USA}
\author{Ziyan Zhu}
\affiliation{Stanford Institute of Materials and Energy Science, SLAC National Accelerator Laboratory, Menlo Park, CA 94025, USA}
\affiliation{Department of Physics, Harvard University, Cambridge, Massachusetts 02138, USA}

\author{Rachel Kurchin}
\affiliation{Department of Materials Science and Engineering, Carnegie Mellon University, Pittsburgh, Pennsylvania 15213, USA}
\author{Efthimios Kaxiras}
\affiliation{Department of Physics, Harvard University, Cambridge, Massachusetts 02138, USA}
\author{Venkatasubramanian Viswanathan}
\affiliation{Department of Mechanical Engineering, Carnegie Mellon University, Pittsburgh, Pennsylvania 15213, USA}
\affiliation{Department of Materials Science and Engineering, Carnegie Mellon University, Pittsburgh, Pennsylvania 15213, USA}
\email{venkvis@cmu.edu}





\begin{abstract}
In this work, we develop a twist-dependent electrochemical activity map, combining a tight-binding electronic structure model with modified Marcus-Hush-Chidsey kinetics in trilayer graphene. We identify a counterintuitive rate enhancement region spanning the magic angle curve and incommensurate twists of the system geometry. We find a broad activity peak with a ruthenium hexamine redox couple in regions corresponding to both magic angles and incommensurate angles, a result qualitatively distinct from the twisted bilayer case. Flat bands and incommensurability offer new avenues for reaction rate enhancements in electrochemical transformations.




\end{abstract}

\maketitle


Enhancing electrochemical reaction rates is critical for chemical transformations \cite{seh2017combining}, energy conversion, and storage \cite{kerman2017practical}. Electrochemical reactions at electrode-electrolyte interfaces are often controlled by modifying the substrate and thereby tuning the adsorption energy to reach peaks of activity ``volcanoes'' \cite{norskov2011density}. Another less-studied approach to enhancing electrode-electrolyte reactions is through modifying the electronic density of states (DoS) of the electrode such that larger numbers of states coincide with the transitions of the redox couple \cite{royea2006comparison,kurchin2020marcus}. For instance, it has been shown that in single-molecule reactions with metallic electrodes (such as gold or copper), rates can increase with increasing surface DoS near the Fermi level \cite{boyen2006local, gu2022increased}.

Interfacial chemical reactivity of few-layer graphene is significant for nanoscale electrochemical devices and electrocatalysis \cite{jin2018emerging, morell2013electronic, craciun2009trilayer}. It offers the possibility of controllable enhancement of the DoS by modifying the twist angle \cite{carr2017twistronics, carr2020electronic}. In a recent work, we showed that twisted bilayer graphene (tBLG) can be used to control the reaction rates with a canonical redox couple, producing an activity volcano of an order of magnitude enhanced rates relative to the untwisted Bernal stacked graphene \cite{yu2022tunable}. However, the reaction rate maximum was confined close to the magic angle (MA) 1.1$^{\circ}$, which would present challenges in fabrication of practical functional devices.

In this work, we analyze the electrochemical activity contour in twisted trilayer graphene (tTLG), which has unique electronic and mechanical relaxation properties associated with incommensurability (i.e. the induced moir\'e cells of the individual pairs of layers cannot be made to coincide as integer multiples) \cite{zhu2020twisted, zhu2020modeling} compared to tBLG. We use a general momentum-space model that incorporates these properties to compute the DoS as a function of twist \cite{zhu2020twisted}. We then map the electron transfer rates given by the overlap of DoS with the redox couple states and electron/hole occupations under an appropriate theory framework, which is the so-called Marcus-Hush-Chidsey model \cite{kurchin2020marcus} (explained in more detail below), modified for the case of low-dimensional electrodes. Finally, we modulate temperature and redox couple parameters to adjust enhancement and magnitude shifts in the activity volcano. 

We show that the rate enhancement is not necessarily localized at the MA in twisted trilayer systems, which can allow better flexibility over a range of electrochemical and redox couple parameters. We find that the rates can increase at angles away from the MA where non-overlapping bands from incommensurability increase the number of interacting states at higher energies. We can ``trade off'' the rate enhancement at these two regions by varying the temperature, where both features become equally significant near the standard condition, forming an activity volcano. We find retention of activity at optimal incommensurate angles over MA at higher redox couple potentials. Our analysis shows that low reorganization energy and high temperature augments the participating solvent states and improves the reaction rate magnitude.

A general scale for electrochemical reaction rates is given by the heterogeneous rate constant ($k_0$), typically computed via the Butler-Volmer (BV) equation \cite{doyle1993modeling,dickinson2020butler}. At higher rates, the linearity assumption in BV breaks down and Marcus theory is needed, which utilizes a second-order free energy expansion \cite{miao_electro-oxidation_2021, boyle2020transient, saveant1975convolution, krauskopf2020physicochemical, dickinson2020butler}. Chidsey introduced a further modification to account for electron and hole occupation functions, known today as the Marcus-Hush-Chidsey (MHC) model \cite{marcus1956theory, chidsey1991free}. The MHC model assumes a constant, energy-independent density of states. This assumption is relaxed in its extended versions \cite{kurchin2020marcus, gerischer1991electron} for better predictions at high overpotentials \cite{narayanan2017dimensionality} and applications like reactivity of graphene edge states and defects \cite{kislenko2020influence, pavlov2020fast} or lithium stripping and electrodeposition \cite{kurchin2020marcus}.  

For materials with sharply varying DoS such as flat-band twisted systems, it is crucial to incorporate DoS effects on the interfacial kinetics with an electrolyte. Due to the large number of states at the magic angle, twisted graphene has the potential for exceptionally high rate enhancement comparable to bulk graphite \cite{yu2022tunable}. We propose that a significant enhancement of the reaction rate takes place when the redox couple energy levels line up with the energy of electrode states whose density has been increased by the tTLG.

\begin{figure}[hbt!]
\includegraphics[width=\linewidth]{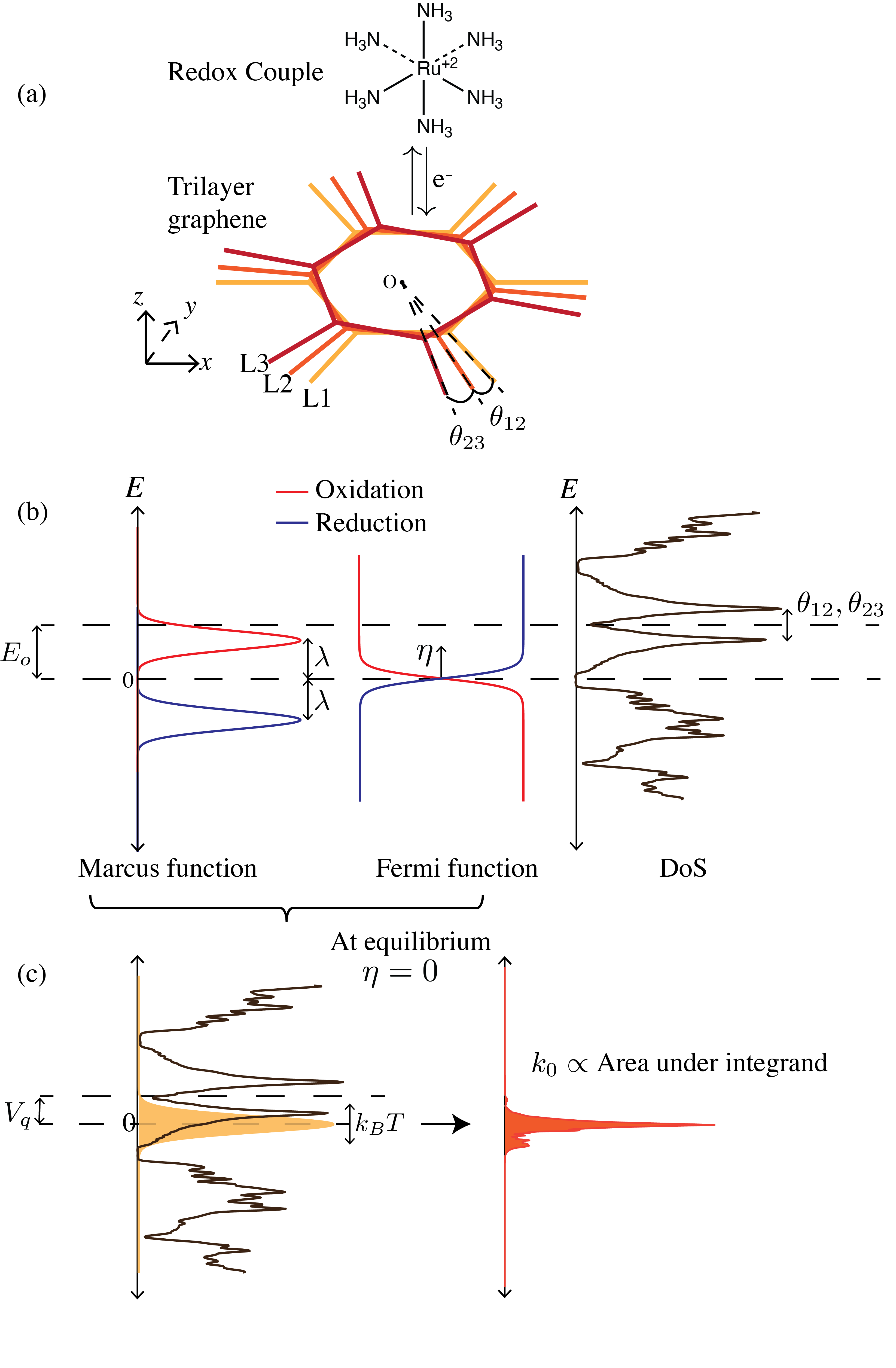}
\caption{\label{fig:schem} Schematic of the reaction system (a) with the trilayer graphene twisted by $\theta_{12}, \theta_{23}$, exchanging electron from a redox couple (RuHex) approaching from $z$. The extended Marcus-Hush-Chidsey model (b) has three terms in the integrand; the Marcus term, Fermi function, and the DoS of the electrode as shown from left to right. All tunable parameters are labelled in (b). Marcus term and the Fermi function illustrated with the same color in (b) multiply to give either oxidation (red) or reduction (blue) filters, identical at equilibrium as indicated in yellow (c). The filter overlaps with the DoS (black) shifted by quantum capacitance voltage $V_q$. The equilibrium rate is proportional to the area under the product of the three terms.}
\end{figure}

We study the electron transfer dynamics between twisted trilayer graphene in contact with a redox couple ($\mathcal{R}$),
\begin{equation}
\ce{\text{tTLG} + \mathcal{R} <=> \text{tTLG}^{+} + \mathcal{R^{-}}}
\label{eq:one}
\end{equation}
approaching from the $z$ direction, as illustrated in the schematic (Fig.~\ref{fig:schem}a). We use ruthenium hexamine (\ce{Ru^{+3}(NH3)6}, RuHex hereafter) in \ce{KCl} solvent as our reference redox couple whose formal potential is closest to the charge neutrality point (CNP) of layered graphene \cite{yu2022tunable}. 

The extended MHC theory \cite{kurchin2020marcus, kislenko2020influence} predicts the reaction rate given the reorganization energy ($\lambda$), applied overpotential ($\eta$), temperature ($T$) and DoS ($\mathcal{D}(\epsilon)$) of the electrode,

\begin{widetext}
\begin{eqnarray}
k_{\text{ox}} &\propto& \frac{1}{\sqrt{4 \pi \lambda k_{\text B}T}}\int \exp \left( {\frac{-(\epsilon + \eta + \lambda)^{2}}{4 \lambda k_{\text B}T}} \right) f(\epsilon, T) \mathcal{D}(\epsilon + eV_{\text{q}}) \text{d}\epsilon, \nonumber 
\\
k_{\text{red}} &\propto& \frac{1}{\sqrt{4 \pi \lambda k_{\text B}T}}\int \exp \left( {\frac{-(\epsilon + \eta - \lambda)^{2}}{4 \lambda k_{\text B}T}} \right) \big(1-f(\epsilon, T)\big) \mathcal{D}(\epsilon + eV_{\text{q}}) \text{d}\epsilon,
\label{eq:two}
\end{eqnarray}
\end{widetext}

\noindent
where $f$ is the Fermi-Dirac distribution. The integrand has three terms: redox couple states (``Marcus-like'' terms), electron or hole occupation probabilities, and the DoS (illustrated in Fig.~\ref{fig:schem}(b)). Assuming area-normalized rates, the proportionality constant is the squared energy-independent electronic coupling term between the electrode and the redox couple \cite{pavlov2020fast, gu2022increased}. Calculation of electronic coupling would require an electronic structure of the ground-state atomic configuration, which is not feasible for the large moir\'e supercells characteristic of small-angle twisted multilayer graphene. 
 
In low-dimensional electrodes with energy dependent DoS \cite{wu2022understanding}, we account for the relatively small and potential-dependent quantum capacitance $C_{\text q}$, which creates a dynamic doping effect \cite{guell2015redox, kislenko2020influence}, by shifting the argument of the DoS by an energy of $e V_{\text{q}}$ as indicated in Eq.~\ref{eq:two}. In bilayer and trilayer graphene, the dielectric capacitance is high enough that this effect can be neglected \cite{zhan2016contribution}, as the voltage drops mainly across quantum and the electrical double layer (EDL) capacitance. At equilibrium, $eV_{\text{q}}$ subtracts from the formal energetic difference between the redox couple and electrode potentials ($E_{0}$), while the rest of the voltage drops across the EDL, $V_{\text{dl}} = E_{0} - V_{\text{q}}$. Assuming the rigid-band approximation, these quantities are obtained from the following coupled system of equations \cite{yang2015density, paek2012computational, yu2022tunable},
\begin{eqnarray}
\Delta E = E_0 + \eta &=& e(V_{\text{dl}} + V_{\text{q}}),
\nonumber \\ 
C_{\text{dl}}V_{\text{dl}} &=& C_{\text{q}}V_{\text q},
\nonumber \\
C_{\text q} &=& e^{2} \int \mathcal{D}(\epsilon) F_\text{t}(\epsilon - eV_{\text q}) \text{d}\epsilon,
\label{eq:three}
\end{eqnarray}
where $\Delta E$ is the difference in the Fermi levels,  $F_\text{t}$ (Eq.~\ref{eq:four}) is the thermal broadening function, and $C_{\text q}$ and $C_{\text{dl}}$ are aggregated quantum and EDL capacitances, respectively. Previous calculations on tBLG \cite{yu2022tunable} suggest that a greater fraction of $E_{0}$ goes into into $V_{\text{q}}$ near the CNP from flat electronic bands. This contributes to a small offset in the rate enhancement away from the MA, also observed in tBLG \cite{yu2022tunable}. 

\begin{equation}
    F_\text{t} = (4k_{\text B}T)^{-1}\text{sech}^{2}(\epsilon/2k_{\text B}T)
\label{eq:four}
\end{equation}

The product of the first two terms (redox couple states and Fermi function) in the integrand of Eq.\ref{eq:two} almost approximates a Gaussian filter being convolved with the DoS. As illustrated in Fig.~\ref{fig:schem}c, when this filter is closer to the flat-band energy, the reaction rate is higher due to a greater overlap with the DoS. A number of control parameters can be tuned that change the shape and position of the overlapping terms in the integrand. Fig.~\ref{fig:schem}(b) highlights the tunable parameters of the solvent ($E_0$, $\lambda$) and the electrode ($\theta_{12}$, $\theta_{23}$, $\eta$). The temperature ($T$), broadens both the redox couple states and the Fermi function.  

Using this model, we have shown a qualitative agreement with experiments in twisted bilayer graphene (tBLG) and RuHex, where the rate enhancement is observed near the MA (1.1$^{\circ}$) \cite{yu2022tunable}. Notably, the rate enhancement is an order of magnitude ($\sim10\times$) higher than the theoretical prediction, possibly due to the twist dependence of the electron coupling constant not included in the model. Modifications to the coupling constant like those studied on graphene edges \cite{unwin2016nanoscale, pavlov2019role} and metallic electrodes \cite{ko2010superior, xie2019determination} can help narrow the gap between theory and experiment.

\begin{figure}[!hbt]
\includegraphics[width=2.6in]{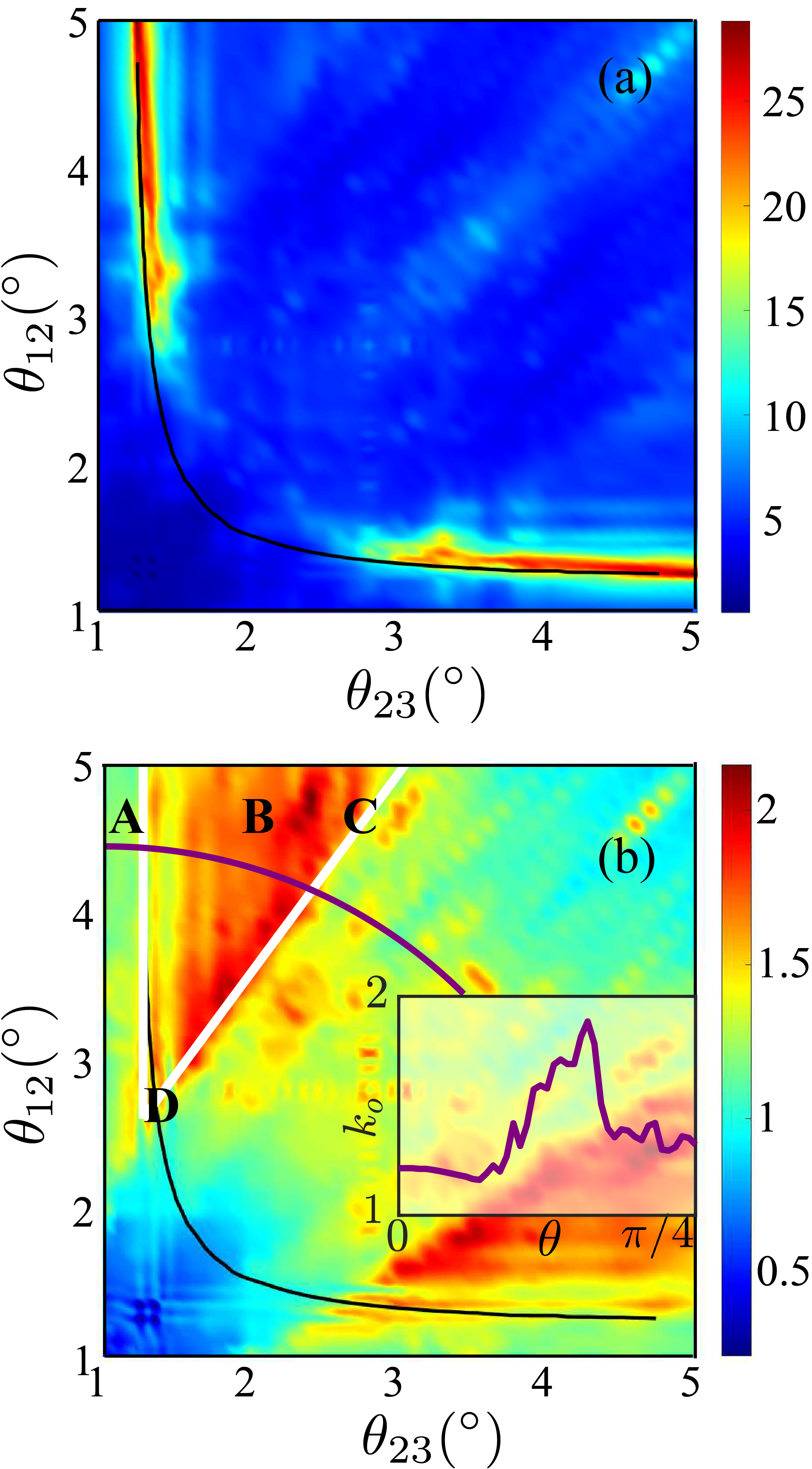}
\caption{\label{fig:komap} Maximum DoS peak (reproduced from \citet{zhu2020twisted}) (a). (b) $k_0$ rate map for RuHex ($E_0= -0.07$ eV, $\lambda = 0.82$ eV, $\eta = 0$ eV) as a function of the two independent twist angles in tTLG system. The ``magic angle curve" is shown by the black line. A number of weak hotspots in $k_0$ (b) occur near the diagonal due to numerical artifacts in the DoS. The solid white triangle encloses the activity volcano, also marked by $\textbf{A}, \textbf{B}, \textbf{C}$ and $\textbf{D}$. Inset shows variation of $k_{0}$ across a 45$^{\circ}$ arc as indicated. All rates are relative to the untwisted ABA system. Maximum rate enhancement in the volcano is $2.1\times$ over ABA.}
\end{figure} 

In the trilayer system, we employ a low-energy momentum-space continuum model for electronic structure and DoS calculations \cite{zhu2020twisted}. The model includes out-of-plane relaxation, which is known to affect the DoS more than the in-plane components. Additionally, when the twist angles are both $\ge1^{\circ}$, the relaxation magnitude is small and the quantitative difference between \cite{carr2018relaxation, carr2019exact}. The model utilizes one twist angle between each adjacent pair of layers ($\theta_{12}$ and $\theta_{23}$, measured in opposite senses but both defined to be $>0$, see Fig.~\ref{fig:schem}a).

The degree of twist modulates the van Hove singularity (VHS) separation and associated DoS peaks (Fig.~ \ref{fig:schem}b). Fig.~\ref{fig:komap}a shows the magnitude of VHS DoS peaks at a range of twist angles (reproduced from \citet{zhu2020twisted}). The high-magnitude peaks asymptotically converge to tBLG MA (1.1$^{\circ}$) when the other layer is decoupled (twisted at a large angle $>$4$^{\circ}$). Plugging the tight-binding DoS into the MHC model, the reaction rate contour ($k_0$) for RuHex (Fig.~\ref{fig:komap}b) contrasts starkly with the DoS peak contour. The color magnitudes are shown relative to the reference ABA system with dispersive bands for direct visualization of the rate enhancement. The comparison assumes no change in electronic coupling with the redox couple between twisted and untwisted systems. Unlike tBLG, the rate enhancement is not confined to the MA curve, but spans a sizeable triangular area (volcano) as indicated in Fig.~\ref{fig:komap}b.  

\begin{figure}[!hbt]
\includegraphics[width=2.8in]{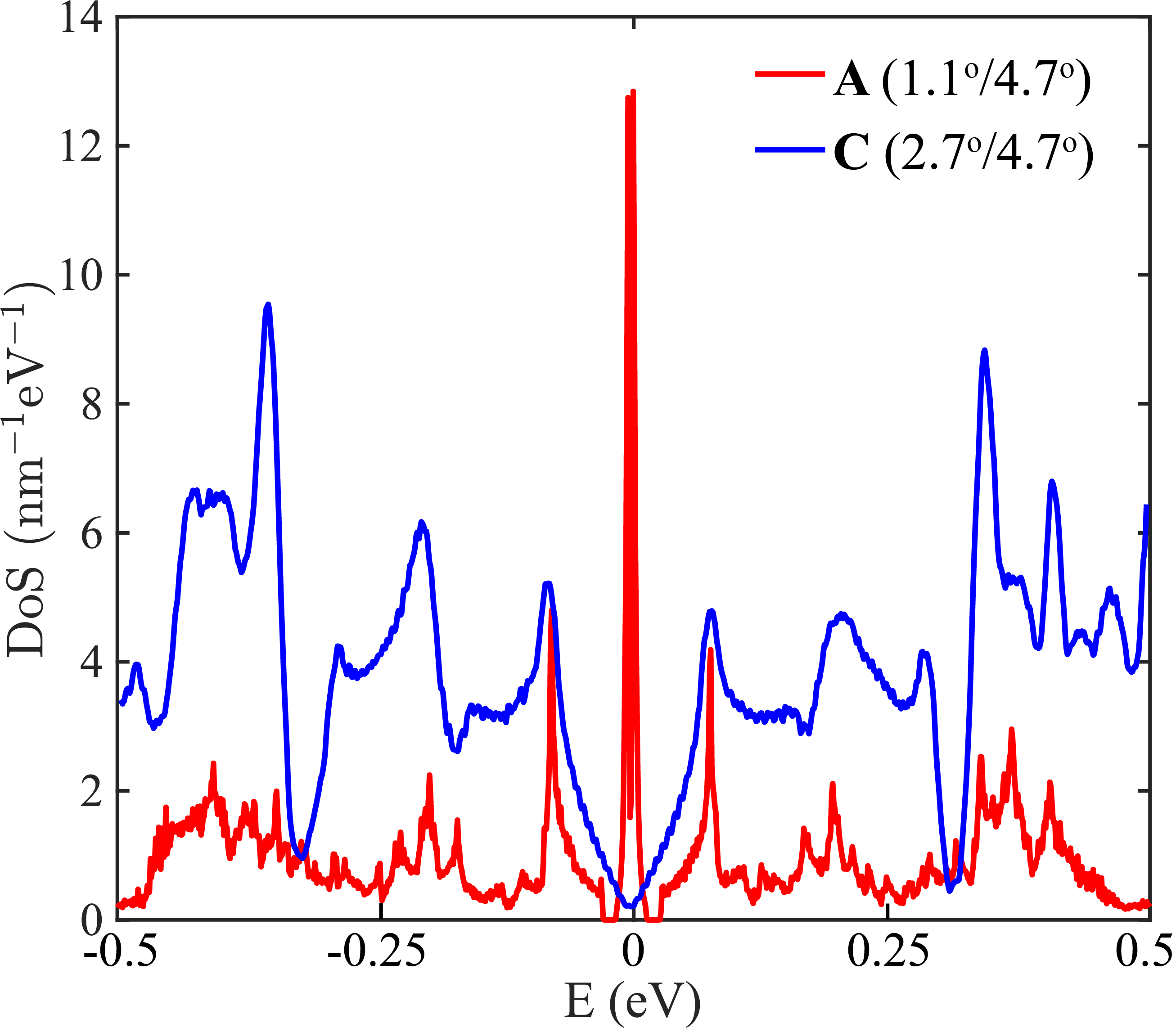}
\caption{\label{fig: ttlg_DoS} Density of states of tTLG at the magic angle \textbf{A} (1.1$^{\circ}$/4.7$^{\circ}$) and at the incommensurate angle \textbf{C} (2.7$^{\circ}$/4.7$^{\circ}$). \textbf{C} has more states at higher energies ($\pm0.25$eV) from non-overlapping bands.}
\end{figure}

In this region, the value of $k_0$ is $\sim2.1\times$ higher than untwisted ABA, which is similar to the theoretical enhancement reported for tBLG ($2.2\times$) \cite{yu2022tunable} with respect to the Bernal stacked AB graphene. The magnitude of $k_0$ in the tTLG volcano is $3.2\times$ higher than tBLG at $1.1^{\circ}$, as expected from the flatter DoS of the former (comparing Fig.~\ref{fig: ttlg_DoS} and S4). Experimentally, this could translate to more than an order of magnitude enhancement in tTLG compared to tBLG. We also show the volcano peak across a 45$^{\circ}$ arc passing over the enhancement region (Fig.~\ref{fig:komap}b inset). 

The activity volcano starts at the MA curve (line $\textbf{AD}$, Fig.~\ref{fig:komap}b) where the DoS peaks are the sharpest. Going horizontally from $\textbf{A}$ to $\textbf{C}$, the VHS separate towards higher energies (Fig.~\ref{fig: ttlg_DoS}), similar to tBLG. In tTLG however, there is an additional change in the band hybridization from commensurate to incommensurate angles, as shown in Fig.~1b in \cite{zhu2020twisted}. These changes are frequent from $\textbf{A}$ to $\textbf{B}$, where $\textbf{B}$ is at the dominant (2,1) moir\'e harmonic line (Fig.~1b \cite{zhu2020twisted}). After $\textbf{B}$, there is a significant region of incommensurate angles until the (1,1) harmonic, in which many competing harmonics coexist, and non-overlapping bands occur over different energy ranges. These bands show an increased DoS in the energy range $\pm0.25$ eV (Fig.~\ref{fig: ttlg_DoS}), where the DoS has an optimal overlap with the MHC filter (Fig. S1). Consequently, the DoS area within $\pm0.25$ eV increases from $\textbf{B}$ to $\textbf{C}$ (Fig. S6), producing a high value for $k_{0}$ in this region. After $\textbf{C}$, the incommensurate states fall outside the filter range, whereby $k_{0}$ disappears quickly. These non-overlapping bands are not found in polytypes of tTLG with unbroken commensurability (Fig. S4) like ``monolayer-twist-bilayer'' (M\textit{t}B) and alternating twist (A\textit{t}A), and are expected to enhance rates only near the MA. Compared to tBLG, these polytypes have flatter bands from correlated electronic phases (Fig. S4) and are predicted to show an order of magnitude higher value of $k_{0}$ in the experiments.

Notably, the vertex of the volcano (\textbf{D}) starts at the point where the (2,1) harmonic line meets the MA curve \cite{zhu2020twisted}. Therefore $\Delta \textbf{ABD}$ is enhanced by the flat bands, and $\Delta \textbf{BCD}$ by non-overlapping bands from incommensurate angles. Hence we observe a shift in maximum rate enhancement region from DoS peak ($\textbf{A}$ at low $T$) to DoS area ($\textbf{C}$ at high $T$). Near the standard conditions (250-300 K), the whole area is kinetically enhanced because both factors are equally significant. The dominance of either factor can be switched as the temperature changes.


\begin{figure}[!htb]
\includegraphics[width=2.6in]{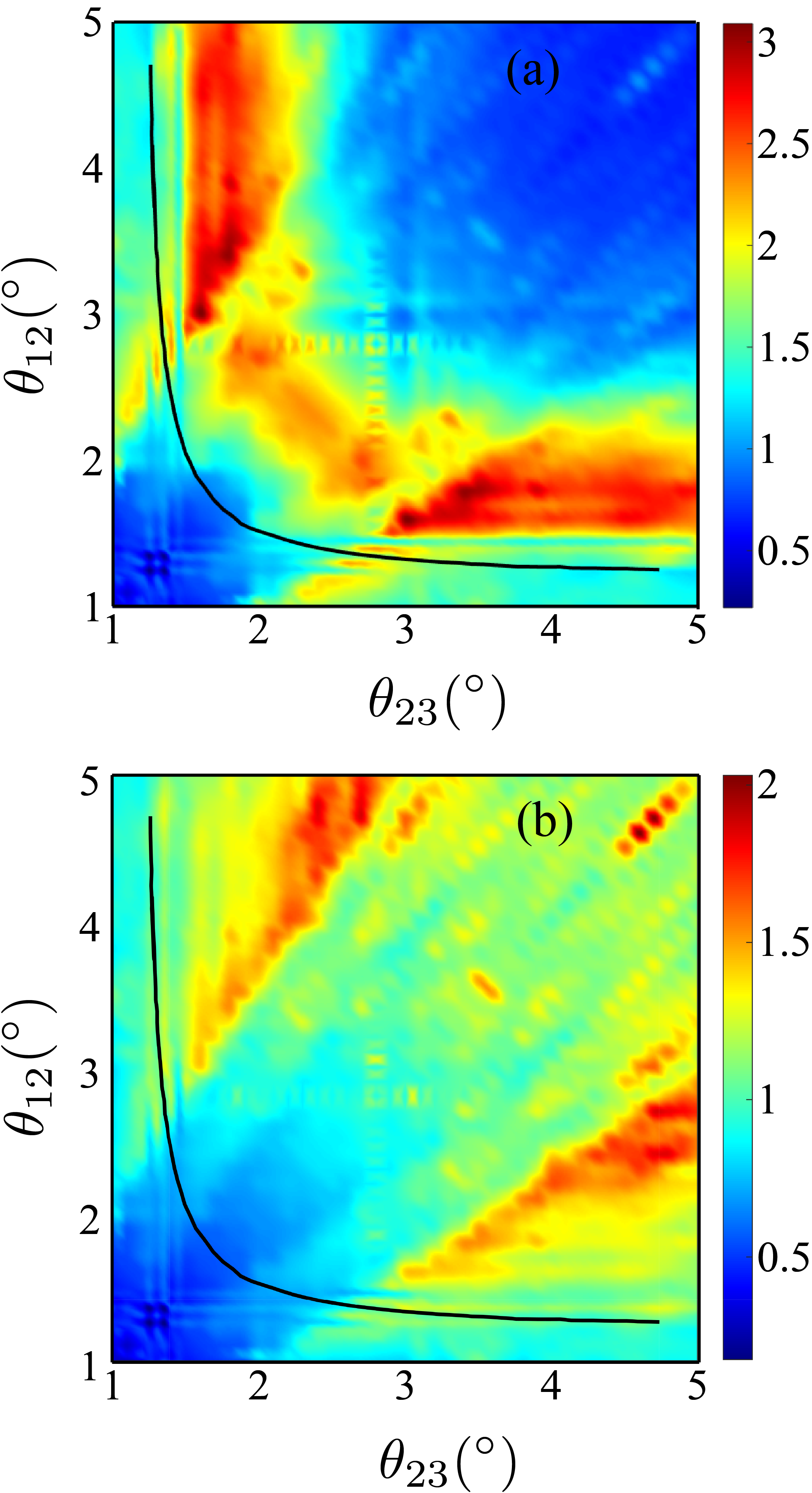}
\caption{\label{fig:T_var} Equilibrium rate constant as a function of temperature; formal potential $E_0 = -0.07$ eV and reorganization energy $\lambda=0.82$ eV set for RuHex solvent at (a) $T = 100$K and (b) $T = 600$K. The solid black line marks the magic angle curve for tTLG. In comparison, the rate enhancement at 100K is higher ($\sim$3) and are closer to the magic angle curve than at 600K. Small filter width at 100K magnifies the effect of flat band singularities, while larger width at 600K promotes higher area under the DoS from non-overlapping bands.}
\end{figure}

Assuming weak $T$ dependence of $\lambda$, DoS and electronic coupling, the absolute magnitude of rates increases rapidly with temperature as the Marcus factor increases. However, there is a shift in the relative rate enhancement from the MA curve, $\Delta \textbf{ABD}$  (Fig.~\ref{fig:T_var}a) to $\Delta \textbf{BCD}$ at higher temperatures (Fig.~\ref{fig:T_var}b). Also, the maximum rate relative to the untwisted system decreases from $3 \times$ at 100 K to $2 \times$ at 600 K. This is due to the small filter width at lower temperatures \cite{pavlov2020fast} (Fig. S1), enabling an efficient overlap with the flat bands at the MA. At higher temperatures the filter overlap with DoS is greater in a higher energy range ($\pm0.25$eV), which is accomplished by non-overlapping bands in $\Delta \textbf{BCD}$ highlighted at 600 K. Note that at 100 K the intensified region is at a small offset from the MA due to the tight overlap with the DoS offset from non-zero quantum capacitance. Hence with a given redox couple, temperature modulations can highlight different features of the DoS in reaction rate contours.


The redox couple parameters ($E_{0}$ and $\lambda$) can be adjusted to study the sensitivity and possible rate improvement. Due to the low energy range of the DoS, we do not show the effect of shifting the filter peak away from $E_{0}=\pm0.3$ eV. Higher values of $E_{0}$ would reduce the overlap with the flat bands as reflected by the gradual shift in the rate enhancement region away from the MA curve at $E_{0} = 0.2$ eV and $0.3$ eV (Fig.~S5). Incommensurate angles, however, retain high activity near \textbf{C} due to the non-overlapping states filling the band gaps. We observe a shift in the rate enhancement to higher incommensurate angles corresponding to the purple region of the dominant moir\'e harmonic plot (Fig.1b of \cite{zhu2020twisted}). Hence, previously inactive redox couples in tBLG like CoPhen (Cobalt Phenanthroline) and FcMeOH (Ferrocene Methanol) \cite{yu2022tunable} ($E_0\sim$0.3 eV) may achieve comparable activity with RuHex at the incommensurate angles.


Tuning the reorganization energy, however, does not produce the same shift in enhancement between MA and the incommensurate region. As shown in Fig.~S2, the filter peak reduces with $\lambda$, but unlike with temperature, there is no effect on the width. Hence, the rate magnitude decreases, but relative to the untwisted system, there is almost no change to the volcano with increasing $\lambda$ (Fig.~S3). These observations suggest the possibility of engineering pairs of redox couples and solvents with low $\lambda$ and high $T$ for high values of $k_{0}$. To boost the effect from flat bands, employ low $E_{0}$ (like RuHex) and $T$ which selectively enhance the twist angles near the MA curve. 

In conclusion, we have explored the role of incommensurability in identifying regions of unexpected rate enhancement in tTLG, which was otherwise locked at the magic angles in commensurate systems like tBLG, M\textit{t}B and A\textit{t}A systems. Incommensurability introduces non-overlapping bands that increase the number of interacting states at higher energies, thus magnifying equilibrium rates away from the MA curve. Temperature variations can adjust rate enhancements between the MA and the incommensurate angles; low temperature contracts the MHC filter thereby favouring flat bands at the MA, and high temperature diffuses the filter favouring greater DoS area at incommensurate angles. With the RuHex redox couple, we observed a high activity volcano spanning both regions at room temperature. Incommensurate angles in twisted trilayer can retain activity with other redox couple like CoPhen and FcMeOH, unlike commensurate systems. The role of twist dependence on the electron coupling constant will be explored in future studies.  This analysis shows the immense flexibility offered by tri-layer systems in maintaining high activity for a wide range of electrochemical processes.

\begin{acknowledgments}
We thank Stephen Carr for important discussions on the tight binding theory. ZZ is supported by a Stanford Science Fellowship. EK acknowledges support through a grant from the Simons Foundation Award no. 896626. Acknowledgment is also made to the Extreme Science and Engineering Discovery Environment (XSEDE) for providing computational resources through Award No. TG-CTS180061.
\end{acknowledgments}


\bibliographystyle{apsrev4-2}
\bibliography{apssamp}

\end{document}


\vspace*{\fill}

\title[suppinfo]{\Large{Supporting Information:\\Twisto-electrochemical activity volcanoes in Trilayer Graphene}}

\author{Mohammad Babar}
\affiliation{Department of Mechanical Engineering, Carnegie Mellon University, Pittsburgh, Pennsylvania 15213, USA}
\author{Ziyan Zhu}
\affiliation{Stanford Institute of Materials and Energy Science, SLAC National Accelerator Laboratory, Menlo Park, CA 94025, USA}
\affiliation{Department of Physics, Harvard University, Cambridge, Massachusetts 02138, USA}

\author{Rachel Kurchin}
\affiliation{Department of Materials Science and Engineering, Carnegie Mellon University, Pittsburgh, Pennsylvania 15213, USA}
\author{Efthimios Kaxiras}
\affiliation{Department of Physics, Harvard University, Cambridge, Massachusetts 02138, USA}
\author{Venkatasubramanian Viswanathan}
\affiliation{Department of Mechanical Engineering, Carnegie Mellon University, Pittsburgh, Pennsylvania 15213, USA}
\affiliation{Department of Materials Science and Engineering, Carnegie Mellon University, Pittsburgh, Pennsylvania 15213, USA}
\email{venkvis@cmu.edu}
\vspace*{\fill}

\maketitle

\begin{figure*}[hbt!]
\centering
\includegraphics[width=4.0in]{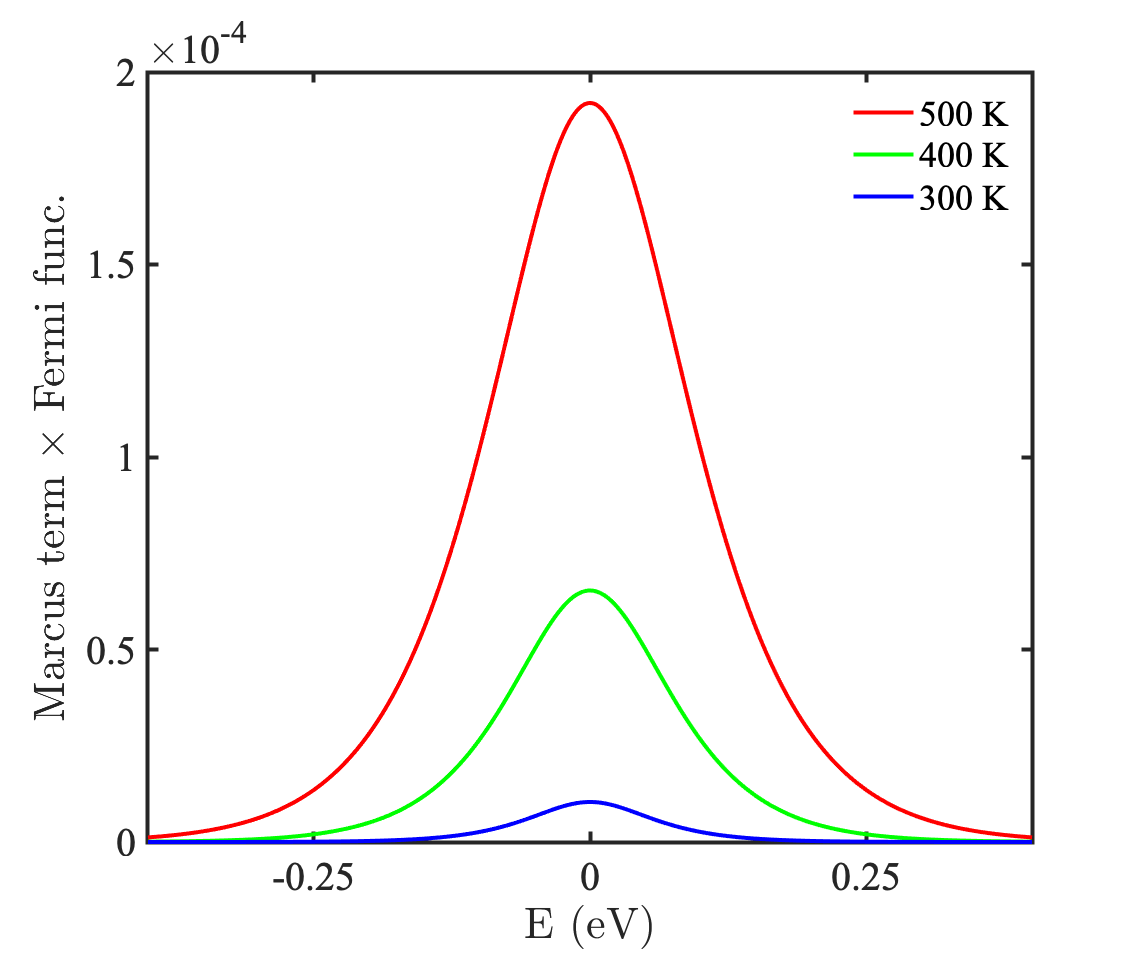}
\caption{\label{fig: T_int} The product of Marcus and the Fermi functions at different temperatures, which acts as an energy filter of the density of states.}
\end{figure*}
\begin{figure*}[hbt!]
\centering
\includegraphics[width=4.0in]{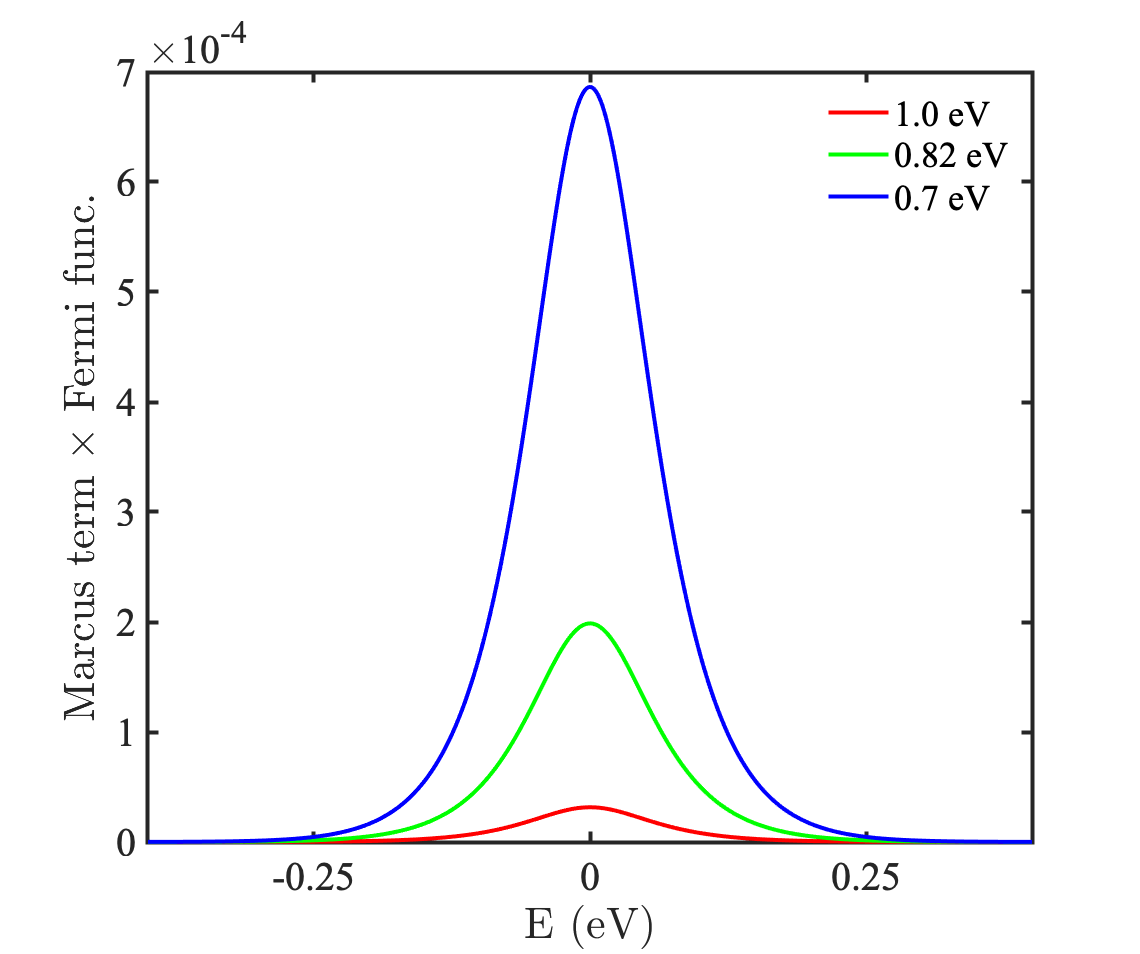}
\caption{\label{fig: l_int} The product of Marcus and the Fermi functions at different reorganization energies.}
\end{figure*}
\begin{figure*}[hbt!]
\centering
\includegraphics[width=2.6in]{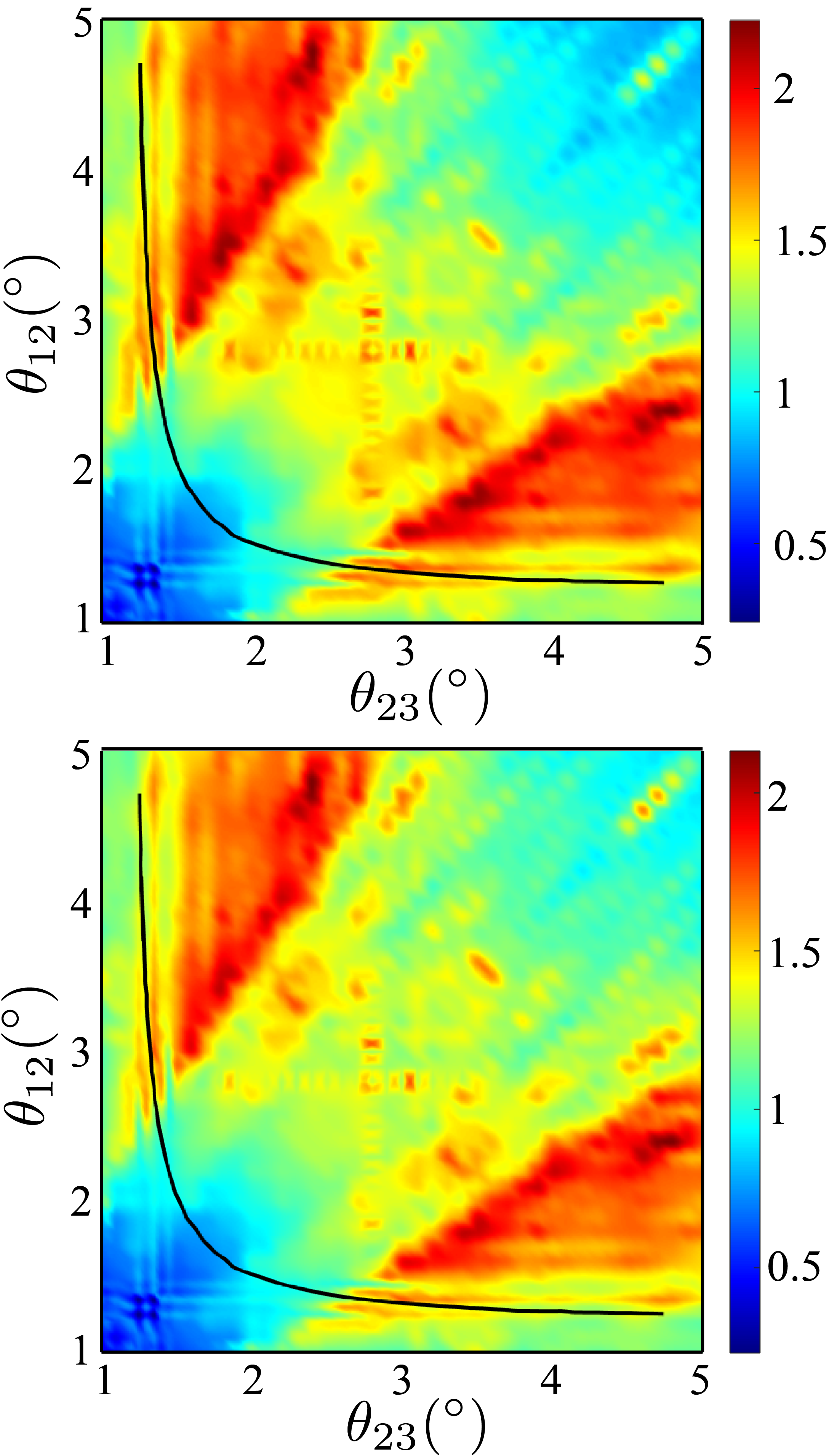}
\caption{\label{fig: l_var} Equilibrium rate maps for (a) $\lambda = 0.2$ eV and (b) $1.2$ eV for RuHex solvent. The rate enhancement region is almost unaffected since the integrand has a constant energy spread with $\lambda$.}
\end{figure*}

The product of the Marcus and Fermi functions inside the rate integrand is shown at different temperatures (fig.~\ref{fig: T_int}) and reorganization energies (fig.~\ref{fig: l_int}). This product overlaps with the density of states to give the overall rate integrand, and is referred to as the MHC filter in the main text. The filter peak and width increases with temperature. The optimal energy range for high overlap of the density of states with the MHC filter is $\pm0.25$eV. Increasing the reorganization energy reduces the peak but the energy spread of the product is not affected. The effect of this is exemplified by the almost identical equilibrium rate maps for two reorganization energies ($\lambda = 0.2$eV, $1.0$eV) at standard values for RuHex ($T$=295K, $E_{0}$=-0.07V, $\eta$=0) as shown in fig.~\ref{fig: l_var}.  

\begin{figure*}[hbt!]
\centering
\includegraphics[width=3.5in]{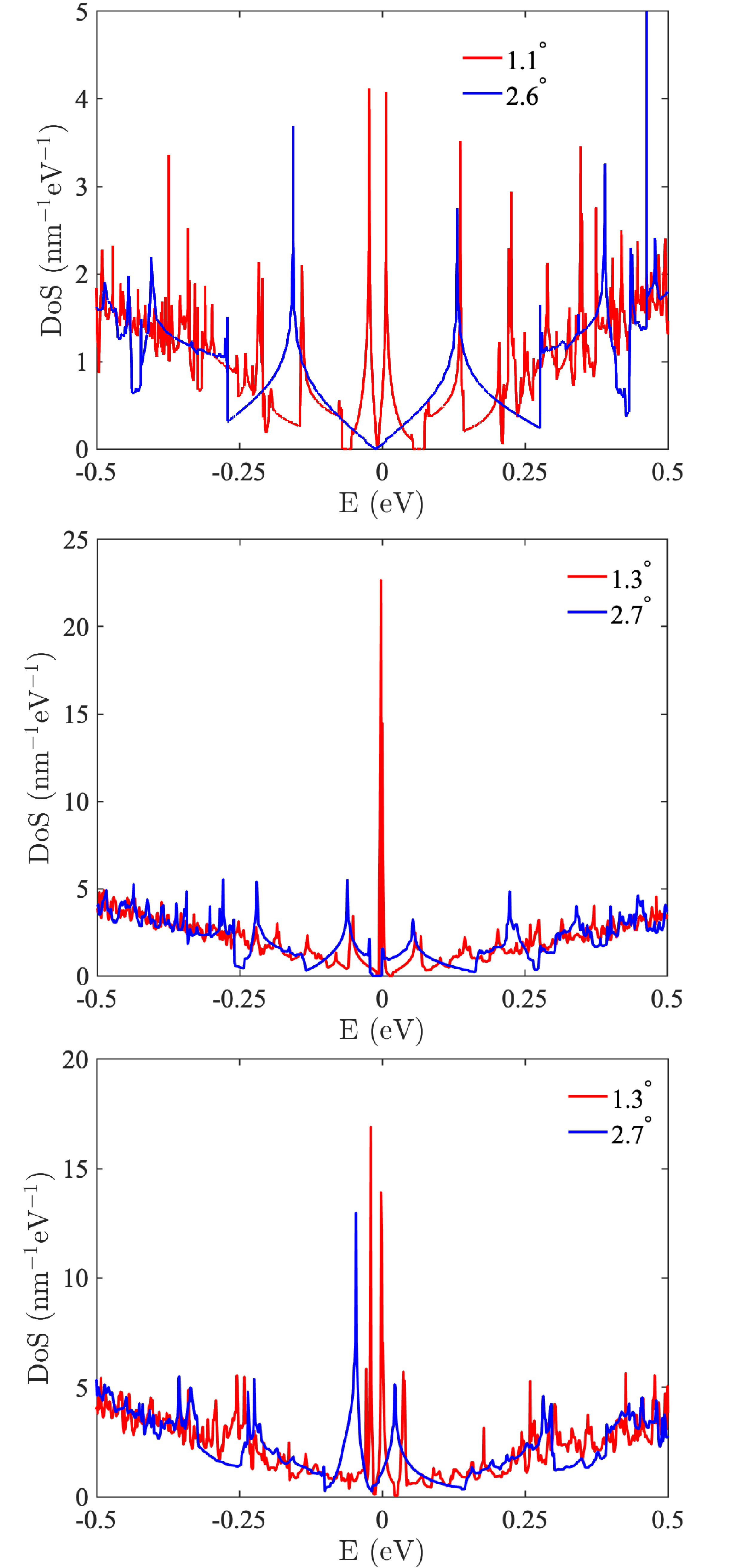}
\caption{\label{fig: tblg_dos} Density of states of commensurate systems near (red) and away (blue) respective magic angles, (a) twisted bilayer graphene (tBLG), (b) monolayer twisted bilayer (M\textit{t}B) and (c) alternating twist trilayer (A\textit{t}A).}
\end{figure*}

The DoS peaks are flatter in tTLG than in tBLG (fig.~3 main text vs fig.~\ref{fig: tblg_dos}(a)), the equilibrium rate therefore is higher ($\times 2$) in the former system at its MA. Like tBLG, polytypes of tTLG, M\textit{t}B and A\textit{t}A are commensurate at all twist angles, and do not show filled band gaps from non-overlapping states at higher energies ($\sim \pm 0.25$ eV) away from their MA (fig.~3 main text vs fig.~\ref{fig: tblg_dos}). As a result, maximum activity in commensurate systems is localized at the MA. 

\begin{figure*}[hbt!]
\centering
\includegraphics[width=2.6in]{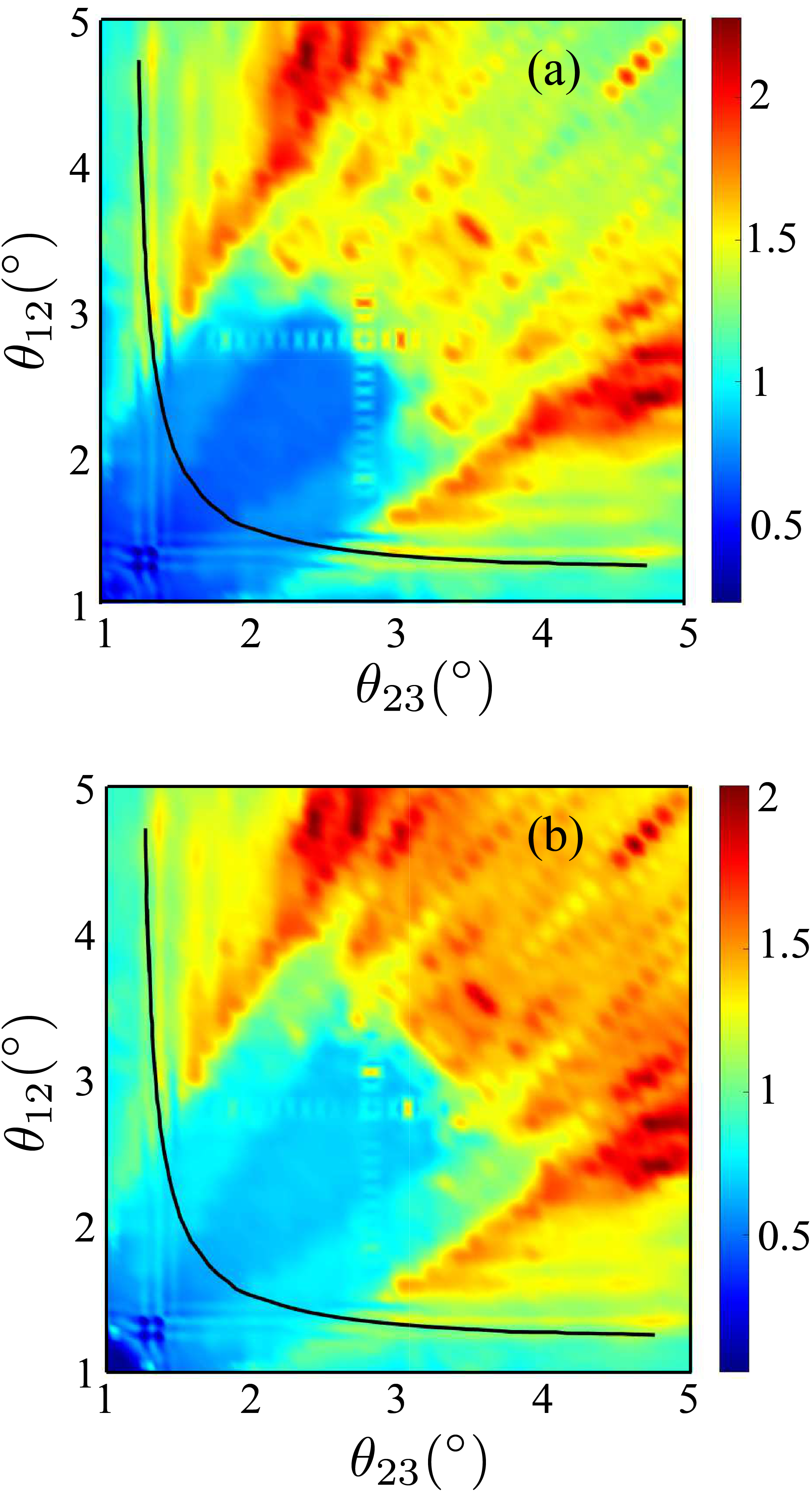}
\caption{\label{fig: eo_var} Equilibrium rate maps for (a) $E_{0} = 0.2$ eV and (b) $0.3$ eV at $\eta = 0$eV, $T = 295$K and $\lambda=0.82$eV. The rate enhancement transfers to higher incommensurate angles, which corresponds to the purple region of the dominant moir\'e harmonic plot shown in fig.1b by \citet{zhu2020twisted}.}
\end{figure*}

Comparison of equilibrium rate maps at higher values of $E_{0}$ is shown in fig.~\ref{fig: eo_var}. As the magnitude of $E_{0}$ increases, the overlap with the flat bands decreases, which shifts the rate enhancement region to higher incommensurate twist angles, while retaining activity at the incommensurate angles (fig.~\ref{fig: eo_var}). The rate enhancement region covers the incommensurate angles marked by the purple region of the dominant moir\'e harmonic plot shown in fig.1b in \cite{zhu2020twisted}.

\begin{figure*}[h!]
\centering
\includegraphics[width=3.3in]{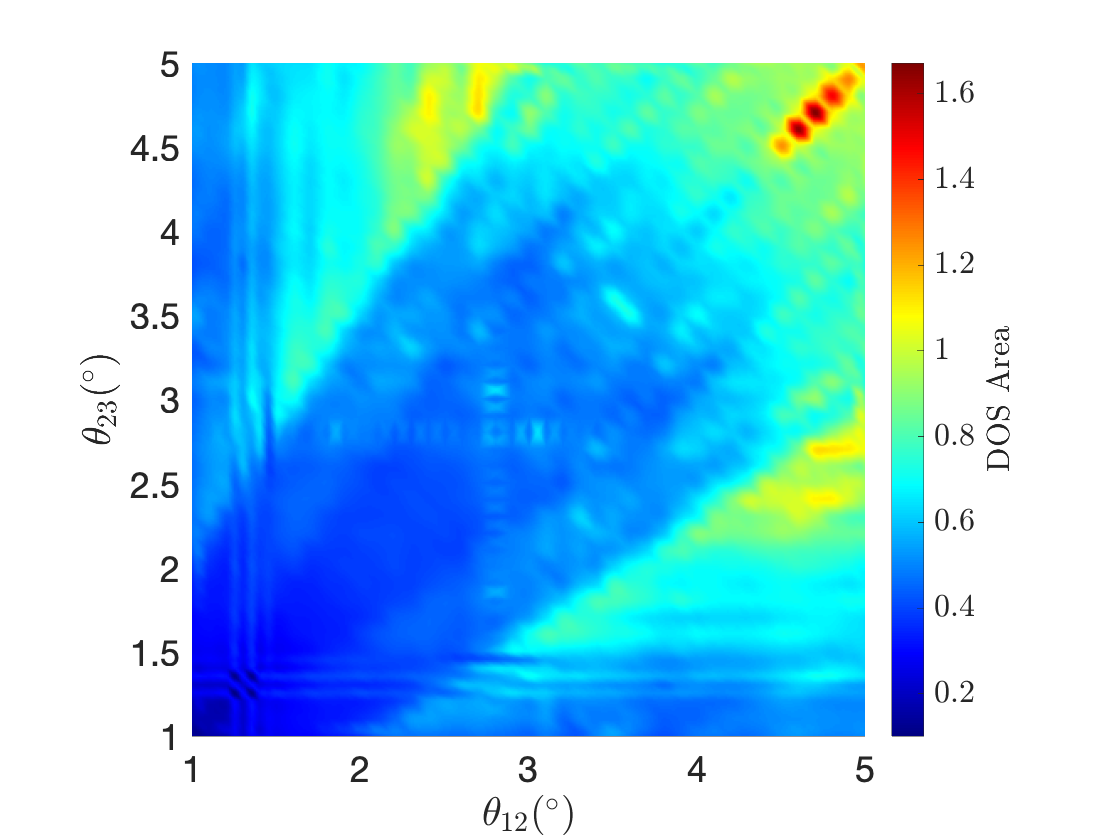}
\caption{\label{fig: dos_norm} Integrated tTLG DoS in $\pm0.25$ eV as a function of twist angles ($\theta_{12}$ and $\theta_{23}$). The DoS area is highest around \textbf{C} ($2.7^{\circ}/-4.7^{\circ}$) within the integration limits. Intense red spots near $\theta_{12}, \theta_{23} = 5^{\circ}$ are due to the numerical artifacts in the DoS.}
\end{figure*}

The map of integrated DoS in $\pm0.25$ eV (fig. \ref{fig: dos_norm}) indicates gradually increasing values of DoS area around point \textbf{C} (fig.~2b of the main text). The maxima occurs at ($2.7^{\circ}/-4.7^{\circ}$). As explained in the main text, a combination of factors are responsible for this behaviour, (a) increase in VHS peak with twist angles from expanding Brillioun Zone (b) flips in band hybridization in the twist angle range. The second factor explains the increase in DoS area from $2.2^{\circ}/-4.7^{\circ}$ (commensurate) to $2.7^{\circ}/-4.7^{\circ}$ (incommensurate) angles. The DoS area drops after \textbf{C} as incommensurate states move out of the integration range (filter width $\pm0.25$ eV). 

\bibliographystyle{apsrev4-2}
\bibliography{apssamp}